# SSE Lossless Compression Method for the Text of the Insignificance of the Lines Order

Juncai Xu*[1], Weidong Zhang[1], Qingwen Ren[1], Xin Xie[2], and Zhengyu Yang[2]

*1 College of Mechanics and Materials, Hohai University, Nanjing, China*

*2 Department of Electrical and Computer Engineering, Northeastern University, Boston, MA USA*

**Abstract**: There is a special type of text which the order of the rows makes no difference (e.g., a word list). To compress these special texts, the traditional lossless compression method is not the ideal choice. A new method that can achieve better compression results for this type of texts is proposed. The texts are pre-processed by a method named SSE and are then compressed through the traditional lossless compression method. Comparison shows that an improved compression result is achieved.

*Corresponding Xu Juncai, PhD, E-mail: xujc@hhu.edu.cn

# 1 Introduction

Data compression can be classified into lossless and lossy. Lossy compression denotes significant data compression with the loss of relatively unimportant details; it can be utilized to compress audio files, images, and videos. However, with regard to text compression, lossless compression is the only choice when the aim is to retain all details. Lossless compression is based on the removal of redundant data, so it focuses on the existence of redundant data.

The simplest lossless compression method is Run-Length Encoding (RLE), which involves encoding continuous repeated data into a specific data amount to achieve compression. It is based on byte or a longer data unit, with byte being the most common one. Another common lossless compression method is entropy coding, which was introduced by Shannon, the father of informatics 0The most widely utilized lossless compression method is Huffman coding**Error! Reference source not found.**0, an extension of Arithmetic Coding 0Given that Arithmetic Coding is protected by various patents, Range Coding which is similared to Arithmetic Coding is more popular in open source communities0The compression limit of entropy coding is entropy, which is represented by H/8 according to the binary system. H is entropy (H =$\sum_{i=1}^{n} p_i * log_2(1/p_i)$), and p is the probability of various characters to appear in the text. Entropy coding has no relations with context, so it cannot remove the context redundant in the text. Compression methods that can remove redundant information in the text are called dictionary encoding. The first dictionary encoding is LZ770 L and Z the initials of the two inventors' names and 77 is the year of invention. LZ77 adopts a slide window to dynamically generate an implicit dictionary. This algorithm can effectively remove redundant information in the context and thus achieves good results in compressing repeated data. Therefore, a series of derivate algorithms have been proposed; the well-known ones include DEFLATE/DEFLATE64 and LZMA/ LZMA2. Another version of LZ is called LZ780Unlike LZ77, LZ78 generates an explicit dictionary and then substitutes the dictionary order for the data in the dictionary. Basically, this method is similar to LZ77. The most famous derivate algorithm of LZ78 is LZW0, which yields a better compression ratio after some improvements. RLE is also a simple dictionary algorithm that can remove redundant information in the context. Other compression algorithms of this type include Dynamic Markov Coding (DMC) 0, Context Tree Weight (CTW) 0, Prediction by Partial Matching (PPM) 0, and PAQ0Basically, these algorithms predict through a model or count redundant data in the context and then employ Arithmetic Coding or other means to effectively remove redundant data. Other algorithms can create redundant data through reversible transformation; such algorithms include BWT algorithm0differential encoding0, and MTF algorithm000These reversible transformation algorithms produce good compression results by creating redundant data and then using other compression methods. Extensive effort has been exerted to develop and update lossless compression methods and obtain a better compression ratio; examples include optimizing existing algorithms on the one hand and finding more effective algorithms (e.g., CSE) on the other hand0

A new method called Sort and Set Empty (SSE) is proposed in this paper. The method can be utilized for the compression of special texts in which the order of the lines makes no difference. The compression result can be optimized by preprocessing the text with SSE and then using lossless compression methods.

## 2 Compression and Decompression Method

### 2.1 Terminology

#### 1) Empty Symbol

An Empty Symbol is a symbol utilized to occupy a position. The symbol should be different from any symbol in the text. For example, in a word list, a space can be used as an Empty Symbol. Several continuous Empty Symbols can be represented by a Empty Symbol followed by a number indicating the number of Empty Symbol (RLE). Corresponding processing should be adopted in decompression.

#### 2) Set Empty

Setting empty involves replacing a character with an Empty Symbol. The procedure can run continuously. Setting several continuous characters Empty can be represented by an Empty Symbol followed by a number (RLE).

#### 3) Sort and Set Empty

This process involves Sorting the special text and then Setting Empty the same characters as those in the previous line.

The SSE lossless compression method employed in this study is a preprocessing method for the text in which the order of lines makes no difference. Owing to the insignificance of the order of lines, the text is first sorted; case sensitivity in the sorting process depend on the instructions of the application layer. After sorting, the beginning of each line has a high level of duplication. A large number of characters can then be Set Empty, producing a text of tree structure (Trie). Finally, the common lossless compression method is adopted to compress the text.

Decompression is the reverse process. First, a common lossless decompression method is adopted. Second, the Empty Symbols are replaced with the characters in the same place in the previous line. The order of the lines in the text is of little importance, so no other processing is required after decompression.

The order of the original text is lost after SSE. Hence, this method can only be applied to the special text or sorted general text.

### 2.2 Pseudo Code

```
const Empty = ‘ ’;
fun Compress(Lines)
  Sort(Lines);
  var o = ‘’;
  for each s in Lines do
    var i = 0;
    var l = o.length();
    loop
      exit when i > l or o[i] <> s[i];
      i += 1;
    end loop;
    o = s;
    println(Empty.x(i) & s.substr(i));
  end do;
end fun;

fun Decompress(Lines)
  var o = ‘’;
  for each s in Lines do
    var i = 0;
    loop
      exit when s[i] <> Empty;
      i += 1;
    end loop;
    o = o.substr(len: i) & s.substr(i);
    println(o);
  end do;
end fun;
```

### 2.3 Experiment

The example was obtained from Free Scrabble Dictionary. The following result was obtained using SSE.

| | **Source text** | **Sort and Set Empty** [1,2] |
|---|---|---|
| 1 | aa | aa |
| 2 | aah | --h |
| 3 | aahed | ---ed |
| 4 | aahing | ---ing |
| 5 | aahs | ---s |
| 6 | aal | --l |
| 7 | aalii | ---ii |
| 8 | aaliis | -----s |
| 9 | aals | ---s |
| 10 | aardvark | --rdvark |
| 11 | aardvarks | --------s |
| 12 | aardwolf | ----wolf |
| 13 | aardwolves | -------ves |
| 14 | aargh | ---gh |
| 15 | aarrgh | ---rgh |
| 16 | aarrghh | ------h |
| 17 | aas | --s |
| 18 | aasvogel | ---vogel |
| 19 | aasvogels | --------s |
| 20 | ab | -b |
| 21 | aba | --a |
| 22 | abaca | ---ca |
| 23 | abacas | -----s |
| 24 | abaci | ----i |
| 25 | aback | ----k |
| 26 | abacterial | ----terial |
| 27 | abacus | ----us |
| 28 | abacuses | ------es |
| 29 | abaft | ---ft |
| 30 | abaka | ---ka |
| 31 | abakas | -----s |
| 32 | abalone | ---lone |
| 33 | abalones | -------s |
| ... | ... | ... |

Table 1

1 Using – as Empty Symbols here

2 Not using RLE code for comparison with original text here

## 3 Compression Ratio Evaluation

The compression method was evaluated comparatively based on two aspects (actual compression and theory compression ratios).

### 3.1 Theory Compression Ratio

Suppose that

$\{x_1, \ldots, x_n\}$ is an alphabet (text alphabet, including line break in the paper).

Then

$\{q_1, \ldots, q_n\}$ is the times in the text, m = $\sum_{i=1}^{n} q_i$ is the length of the text, and

$\{p_1, \ldots, p_n\}$ is its probability of appearing in the text; $\sum_{i=1}^{n} p_i$ = 1.

Information entropy is obtained as H=$\sum_{i=1}^{n} p_i * log_2(1/p_i)$, and the compression ratio is H/8.

This type of text is a special text with an independent line sequence. Thus, the SSE (Sort and Set Empty) method can be applied. The above model is described as follows:

Suppose that

$\{x_0, x_1, \ldots, x_n\}$ is a new alphabet (add $x_0$ as a Empty Symbol).

Then

$\{q`_0, q`_1, \ldots, q`_n\}$ is the times in the text, m is the length of the text, and

$\{p`_0, p`_1, \ldots, p`_n\}$ is its probability of appearing in the text; $\sum_{i=0}^{n} p`_i$ = 1.

The information entropy of the new text is obtained as H` = $\sum_{i=0}^{n} p`_i * log_2(1/p`_i)$; the compression ratio is H`/8.

Consequently, the following formulas are obtained.

(1) $q`_1 \le q_1, \ldots, q`_n \le q_n$

(2) $q`_0 = \sum_{i=1}^{n}(q_i - q`_i)$

(3) $0 \le p`_1 \le p_1 \le 1, \ldots, 0 \le p`_n \le p_n \le 1$

(4) $0 \le p`_0 = \sum_{i=1}^{n}(p_i - p`_i) \le 1$

Function $y = x * \log_2(1/x)$ is shown below:

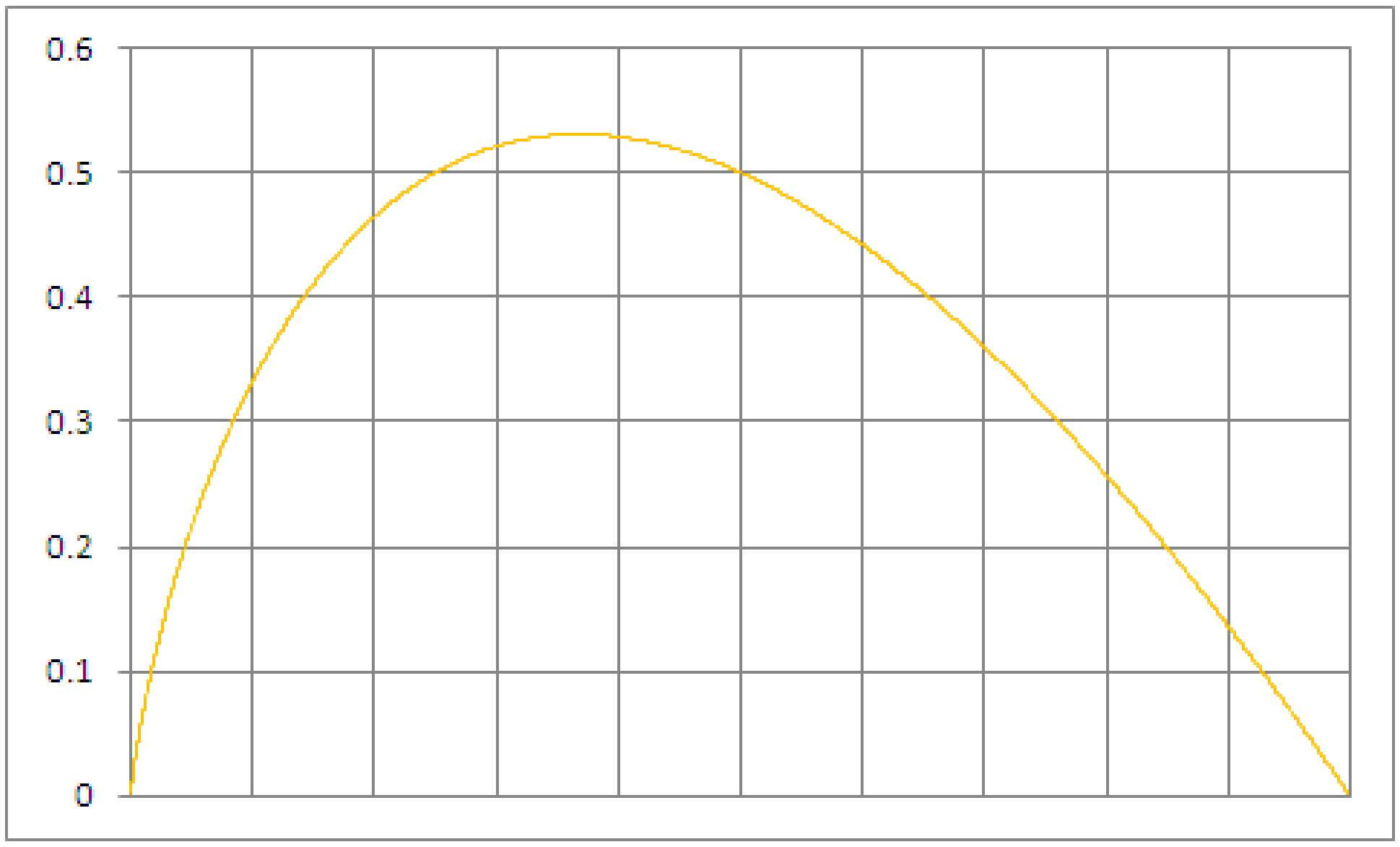

y = x * log2(1/x), x in (0, 1]

Figure 1

The function extreme value is 1/e/log(2) ≅ 0.530737845423043 at 1/e ≅ 0.3678794441171442. The function increases steeply at the left branch of 1/e. The function decreases at the right branch of 1/e. However, the degree of steepness at the right branch is less than that at the left branch.

Compared with H, $H'$ has an additional item, that is, $p'_0 * \log_2 (1/p'_0)$. The maximum of the additional item is 0.53. For every $p'_i \leq p_i$ from 1 to n, the formula is as follows:

(5) $p'_i * \log_2 (1/p'_i) \leq p_i * \log_2 (1/p_i)$, $p_i \leq 1/e$.

(6) If $p_i > 1/e$, the formula on the two sides of (5) is uncertain. The relationship between the two sides maybe greater than, equal to, or less than.

According to the two formulas above, we found that SSE methods can reduce entropy when the probability of the character appearing in the text is less than or equal to 1/e (approximately 0.37). However, when each character probability is larger than 1/e in the text, the SSE methods may increase entropy and result in a large compressed file.

A probability value that is greater than 0.37 (for simplicity, the approximate value is used instead of 1/e) should not exceed 2. The Sort and Set Empty method fails when only two characters exist in the alphabet case.

The Sort and Set Empty method has an effect on the alphabet with more than two characters. The magnitude of the effect has a positive correlation with the following numerical value.

For each alphabet occurrence probability that is less than or equal to 0.37 characters, the SSE is the more times, effect display significant. For a probability value that is less than or equal to 0.37, entropy is reduced. If it is greater than the new Empty Symbol entropy (0.53 maximum), the appearance probability would be greater than 0.37 to increase the total entropy of a few characters. This condition means that the Sort and Set Empty method takes effect and obtains a better compression rate.

Below is an alphabet with a random probability and 2 to 52 characters. In the average entropy statistical graph, "Source" is the average entropy not being processed. "Target" is the average entropy being SSE. The compression effect on the alphabet with number of characters beginning at 5 is significant.

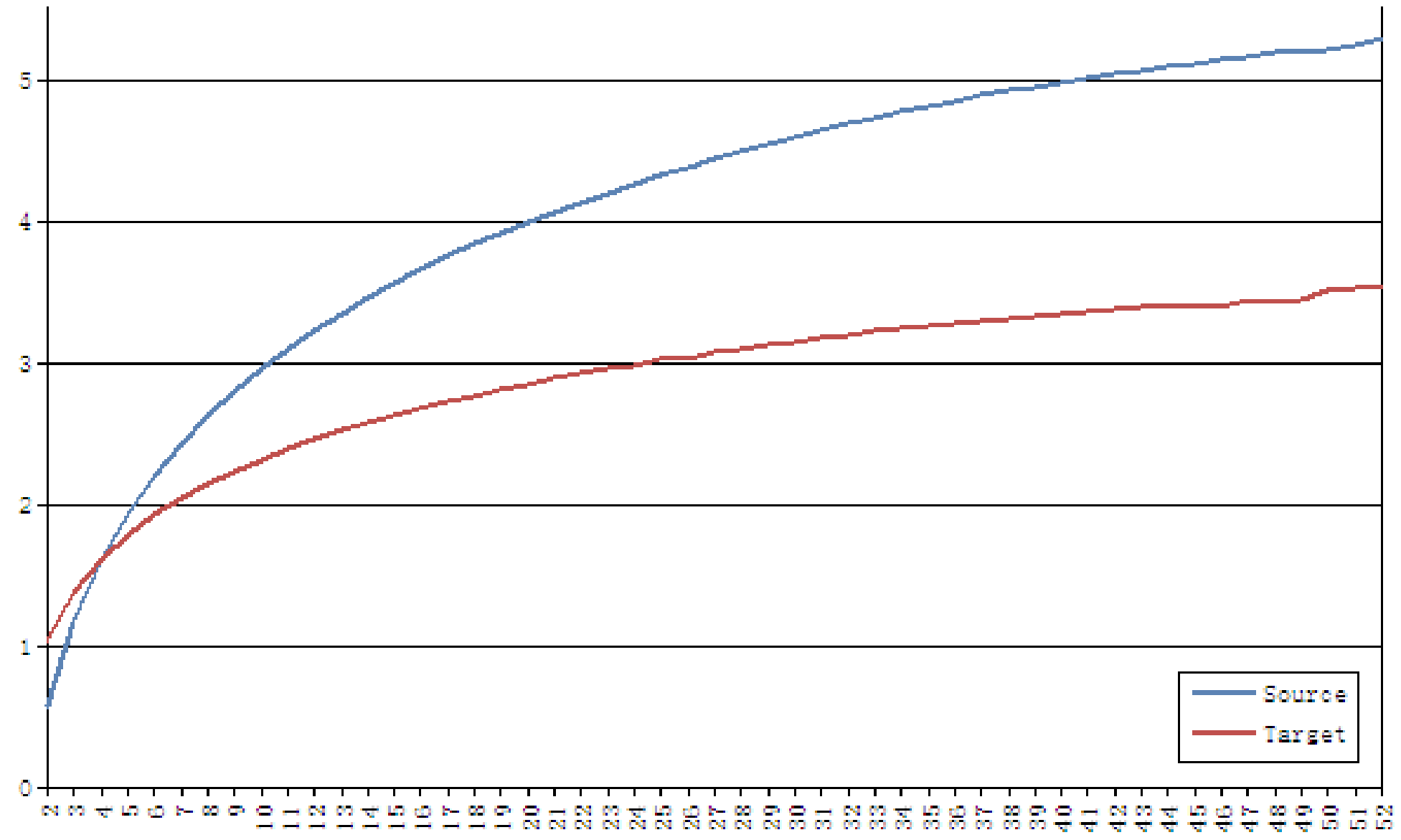

Figure 2

The chart below shows a statistical comparison of average, maximum, and minimum entropy.

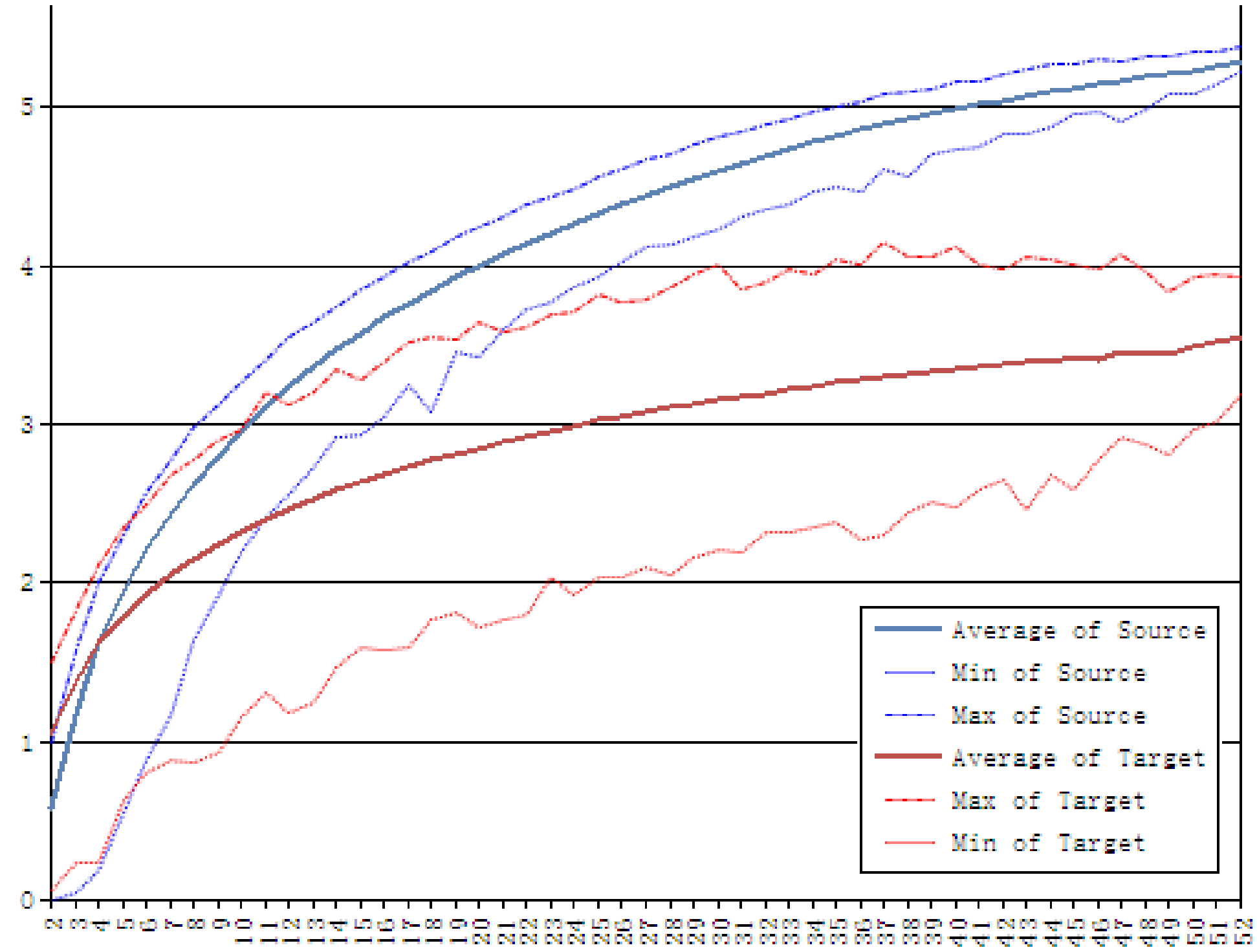

Figure 3

The ratio of average entropy or the average compression ratio (60% to 80%) is shown below.

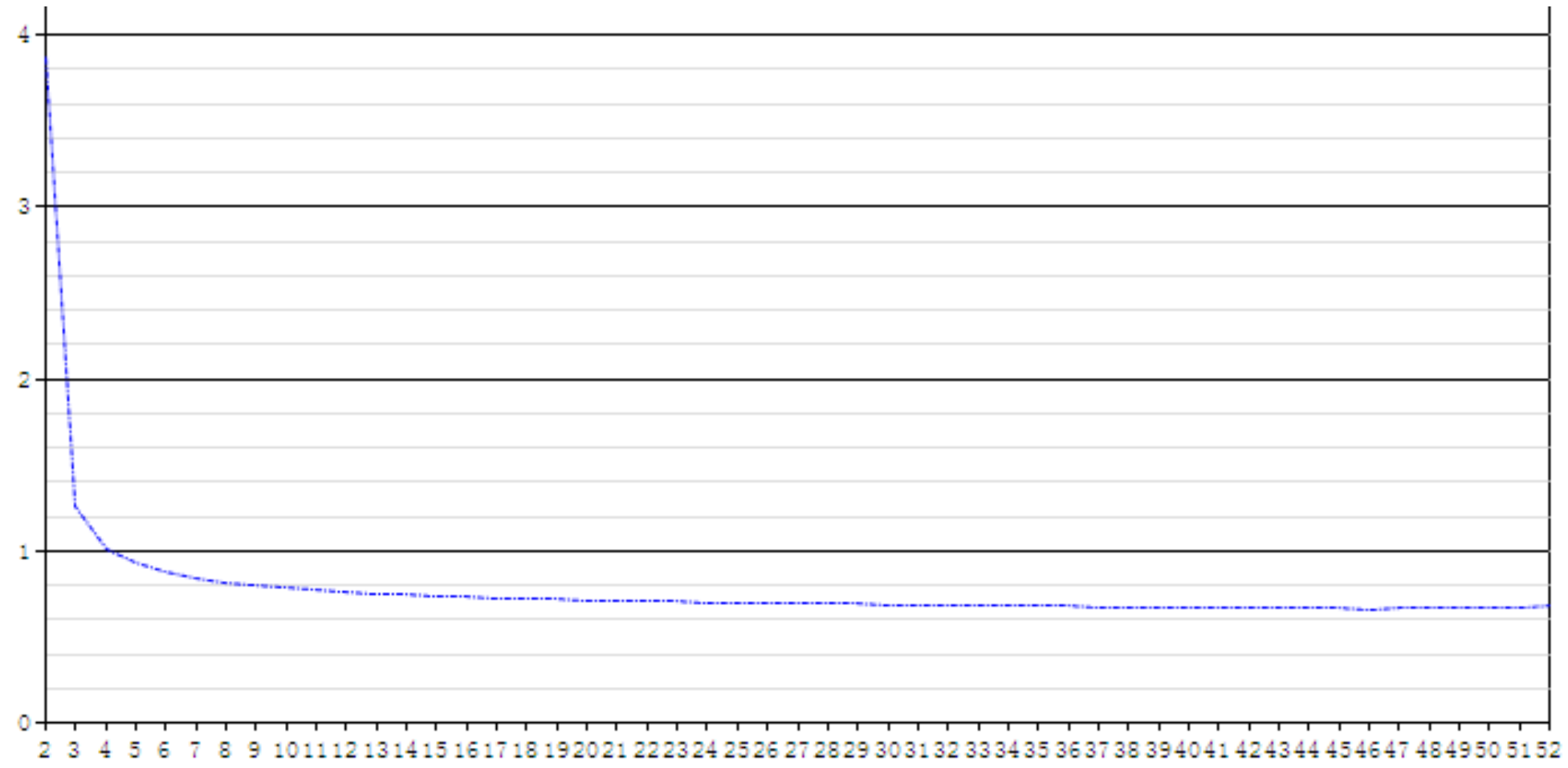

Figure 4

### 3.2 Comparison of compression ratios

Three types of text are presented in the table below for comparison; these three are word list, URL list, and MD5 list. We tested many files for each type and used the average values. We listed two types of compression ratios: entropy and actual compression ratios. The entropy compression ratio employs the binary entropy value. The actual compression ratio employs the Ultra LZMA compression ratio from 7-zip.

Comparison of compression ratios: Source and SSE

| | **Source compression ratio** | | **SSE compression ratio** | | **Ratio of compression ratios1** | |
|---|---|---|---|---|---|---|
| | Entropy2 | Actual3 | Entropy | Actual | Entropy | Actual |
| Word List | 53.64% | 19.45% | 25.92% | 11.37% | 48.32% | **58.46%** |
| URL List | 57.53% | 6.01% | **21.27%** | **4.39%** | **36.98%** | 72.96% |
| MD5 List | 52.12% | 43.50% | 47.93% | 37.89% | 91.95% | 87.10% |

Table 2

The source entropy compression ratio is greater than 50%. All of the ratios are reduced significantly after SSE. In particular, the entropy compression ratio of the URL list is 21.27%. Many repeated prefixes exist in the URL list, so entropy was reduced considerably with SSE. However, fewer repeated prefixes exist in the MD5 list, which has too much random data; hence, entropy was reduced only slightly. The word list is in between.

The actual compression ratios of LZMA are all ideal and are better than the entropy compression ratio (LZMA employs the dictionary compression method, and we used the 7-zip LZMA Ultra option). The best compression ratio of LZMA is for the URL list: the source has 6.01%, and SSE has 4.39%. SSE improvement in the actual compression ratio is weaker than that in the entropy one. For the word list, SSE is ideal; the actual compression ratio is reduced to 11.37% from 19.45%, and the ratio of the ratios is 58.46%. For the MD5 list, the improvement by SSE is the weakest regardless of the actual entropy compression ratio.

1 Ratio of Compress ratio = SSE compress ratio / Source compress ratio

2 Compress ratio = H / 8, H = $\sum_{i=1}^{n} p_i * log_2(1/p_i)$

3 Use 7-zip LZMA Ultra, Compress ratio = Compressed size / source size

## 4 Conclusion

The presented description and actual test showed that the SSE method is useful. For a special text that is row-order independent, SSE can be utilized to preprocess the text and then compress it through other compression methods. The compression ratio would be reduced significantly. SSE is very simple, efficient, and has a good application potential.

Our next work would be to establish a proper method of applying SSE to normal text. Prior to implementing SSE, other methods would be used to remember the positions and restore them after decompression. Such work would extend the application scope of SSE and is worth pursuing.

**Acknowledgement**

This research was funded by the National Natural Science Foundation of China (Grant No. 11132003) and Jiangsu Province post-doctor Foundation of China (Grant No. 1401124C).

Vol.57 (3), pp.311-320OCLC